\documentclass[twocolumn,amsmath,amssymb,prl, lettersize,superscriptaddress,floatfix,footinbib]{revtex4-1}

\usepackage{graphicx}
\usepackage{dcolumn}
\usepackage{bm}
\usepackage{amsmath}
\usepackage{color}

\newcommand{\eqnreft}[1]{{Eq.~(\ref{#1})}}

\newcommand{\eqnsreft}[2]{{Eqs.~(\ref{#1}) and (\ref{#2})}}

\begin{document}

\title{Quantitative acoustic models for superfluid circuits}

\author{Guillaume~Gauthier}
\affiliation{ARC Centre of Excellence for Engineered Quantum Systems, School of Mathematics and Physics, University of Queensland, Brisbane, QLD 4072, Australia}
\author{Stuart~S.~Szigeti}
\affiliation{ARC Centre of Excellence for Engineered Quantum Systems, School of Mathematics and Physics, University of Queensland, Brisbane, QLD 4072, Australia}
\affiliation{Department of Quantum Science, Research School of Physics, The Australian National University, Canberra 2601, Australia}
\author{Matthew~T.~Reeves}
\affiliation{ARC Centre of Excellence in Future Low-Energy Electronics Technologies, School of Mathematics and Physics, University of Queensland, Brisbane, QLD 4072, Australia}
\author{Mark~Baker}
\affiliation{ARC Centre of Excellence for Engineered Quantum Systems, School of Mathematics and Physics, University of Queensland, Brisbane, QLD 4072, Australia}
\author{Thomas~A.~Bell}
\affiliation{ARC Centre of Excellence for Engineered Quantum Systems, School of Mathematics and Physics, University of Queensland, Brisbane, QLD 4072, Australia}
\author{Halina~Rubinsztein-Dunlop}
\affiliation{ARC Centre of Excellence for Engineered Quantum Systems, School of Mathematics and Physics, University of Queensland, Brisbane, QLD 4072, Australia}
\author{Matthew J.~Davis}
\affiliation{ARC Centre of Excellence for Engineered Quantum Systems, School of Mathematics and Physics, University of Queensland, Brisbane, QLD 4072, Australia}
\affiliation{ARC Centre of Excellence in Future Low-Energy Electronics Technologies, School of Mathematics and Physics, University of Queensland, Brisbane, QLD 4072, Australia}
\author{Tyler~W.~Neely}
\affiliation{ARC Centre of Excellence for Engineered Quantum Systems, School of Mathematics and Physics, University of Queensland, Brisbane, QLD 4072, Australia}
\email{t.neely@uq.edu.au}

\begin{abstract}
We experimentally realize a highly tunable superfluid oscillator circuit in a quantum gas of ultracold atoms and develop and verify a simple lumped-element description of this circuit. At low oscillator currents, we demonstrate that the circuit is accurately described as a Helmholtz resonator, a fundamental element of acoustic circuits. At larger currents, the breakdown of the Helmholtz regime is heralded by a turbulent shedding of vortices and density waves. Although a simple phase-slip model offers qualitative insights into the circuit's resistive behavior, our results indicate deviations from the phase-slip model. A full understanding of the dissipation in superfluid circuits will thus require the development of empirical models of the turbulent dynamics in this system, as have been developed for classical acoustic systems.
\end{abstract}

\maketitle

The emerging field of atomtronics aims to build devices using ultracold atomic gases~\cite{wright_driving_2013}, with potential applications including inertial and magnetic sensing~\cite{McDonald:2014, Fang:2012} and quantum simulation~\cite{Gross:2017}. Ultracold atoms also present a configurable and controllable platform for elucidating the foundational principles of superfluid circuitry. Realizing quantitative models for atomtronic circuits will therefore aid the development of other technologies, such as those based on superfluid helium and room-temperature exciton-polariton condensates, where practical applications are forthcoming~\cite{Avenel1997,Hoskinson2006,Sanvitto2016, harris2016laser}.
 
Recent experiments have performed detailed studies of simple atomtronic devices, including the transport of atomic gas superfluids through 
tunnel junctions~\cite{albiez_direct_2005,levy_.c._2007,valtolina_josephson_2015,spagnolli_crossing_2017,burchianti2018connecting}, weak links~\cite{moulder_quantized_2012,wright_driving_2013,jendrzejewski2014resistive,eckel_interferometric_2014}, and mesoscopic channels~\cite{stadler_observing_2012,lee_analogs_2013,eckel_contact_2016}. Although the dynamics of these experiments with Bose-Einstein condensates (BECs) can be accurately modeled using the Gross-Pitaevskii equation (GPE), this is time consuming and is therefore not a feasible approach to atomtronic circuit testing and design. Practical atomtronics will require accurate and quantitative circuit models that include basic elements akin to resistors and capacitors. To a large extent, this has been achieved for superfluid Josephson junctions, where various forms of two-mode models have been successful in modeling real experiments~\cite{spagnolli_crossing_2017}.  However, currently there is no established lumped-element description of superfluid transport through extended channels~\cite{eckel_contact_2016,li_superfluid_2016,lee_analogs_2013}. 

In this Letter, we experimentally study a superfluid system consisting of two reservoirs connected by a configurable linear channel~\cite{stadler_observing_2012,lee_analogs_2013,eckel_contact_2016,li_superfluid_2016}. We perform a comprehensive investigation of this prototypical atomtronic circuit by varying both the channel dimensions and the initial superfluid imbalance between the two reservoirs. We demonstrate that accurate modeling of superfluid atomtronic circuits requires an acoustic description of the device. Unlike circuit models based on electrical analog~~\cite{lee_analogs_2013,eckel_contact_2016}, the acoustic description delivers quantitatively accurate predictions for the circuit oscillation frequency. We also study dissipation in the circuit, and in contrast to recent experiments~\cite{eckel_contact_2016,jendrzejewski2014resistive}, we show that idealized phase-slip models of dissipation fail to accurately describe the circuit's resistive behavior at high initial bias. These results present a new paradigm for developing atomtronic devices, suggesting much can be learned from existing acoustic circuitry principles. Furthermore, an understanding of turbulence and the interactions of excitations will likely be required to completely describe atomtronic circuit resistance.

\begin{figure}[t!]
\includegraphics[width=\linewidth]{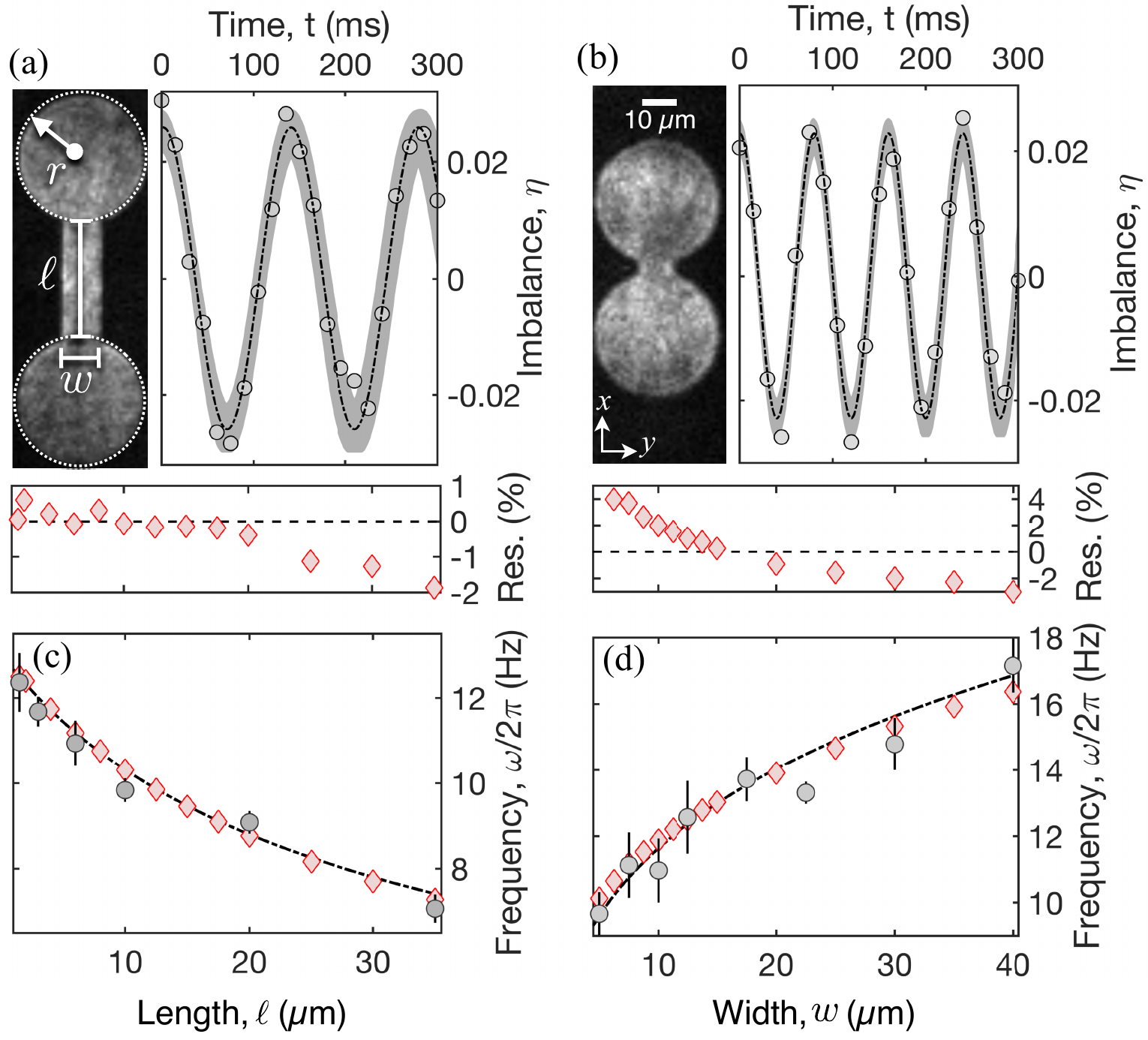}%
\caption{Experimental system and oscillator frequency dependence in the nondissipative regime. {(a)} (Left) \emph{In situ} experimental image for a reservoir radius $r=20$~$\mu$m, channel width $w = 12.5$~$\mu$m, and channel length $\ell=35$~$\mu$m. (Right) Nondissipative dynamics of the population imbalance for this circuit geometry (red points), which are well fitted by a sinusoid (black curve, grey shading indicates 95\% confidence interval). {(b)} As for {(a)} but with $\ell=  1.5$~$\mu$m. {(c)} (Bottom) By measuring oscillations for different atom numbers~\cite{refSupp}, frequencies are extrapolated to $N=2\times 10^6$ atoms (grey circles) for comparison with GPE simulations (orange diamonds) as a function of channel length for a fixed width $w = 12.5$~$\mu$m. The black dash-dot line is a one-parameter fit of the acoustic model to the GPE data, where $\mathcal{V}_1 = \mathcal{V}_2 = \pi r^2 \cdot 2 l_z$ and $\mathcal{S} = w\cdot 2 l_z$, where $l_z$ is the Thomas-Fermi radius in the $z$-direction, resulting in an end correction of $\delta = 2.18(1)\sqrt{\mathcal{S}}$. (Top) Shows the fit residuals. {(d)} Similar to {(c)}, but for varying channel width with a fixed length $\ell = 1.5$~$\mu$m, with the fit giving $\delta = 2.14(3)\sqrt{\mathcal{S}}$. The speed of sound used in the fits was $c = \sqrt{\bar{n} g/m}$, where $\bar{n}$ is the spatially averaged atom number density of the GPE ground state. $c= 2376$~$\mu$m/s and $c = 2332$~$\mu$m/s in {(c)} and {(d)}, respectively.}
\label{fig:Frequency_Dependence}
\end{figure}

\textit{Circuit overview.---} Figure~\ref{fig:Frequency_Dependence} provides an overview of our experiment and its behavior. The BEC is tightly trapped in the vertical direction using an elliptical Gaussian beam, and confined transversely with a hard-walled optical potential resulting from the direct projection of a digital micromirror device (DMD)~\cite{refSupp,gauthier_direct_2016,gauthier2018negative}. An image of the balanced superfluid-filled resonator at equilibrium is shown in Fig.~1(a). By applying a variable linear potential in the $x$ direction~\cite{refSupp}, we can continuously vary the initial population imbalance $\eta = \left(N_{1}-N_{2}\right)/\left(N_{1}+N_{2}\right)$ between the two reservoirs from zero to near 100\%, where $N_{1}$ and $N_{2}$ are measured atom numbers in the top and bottom halves of the circuit, respectively. The sudden removal of the linear potential causes the superfluid to begin to flow, as illustrated in Figs.~\ref{fig:Frequency_Dependence}(a) and \ref{fig:Frequency_Dependence}(b).

\textit{Nondissipative regime.---}  We first examine small initial imbalances  where the superfluid undergoes undamped oscillations between the two reservoirs, similar to a Josephson junction~\footnote{The amplitude of the oscillations is weakly-dependent on the initial bias, increasing with the initial imbalance~\cite{refSupp}.}, for a wide range of channel widths and lengths.  Example results are shown in Figs.~\ref{fig:Frequency_Dependence}(a) and \ref{fig:Frequency_Dependence}(b), where it can be seen that the oscillation frequency decreases for a longer channel. In Fig.~\ref{fig:Frequency_Dependence}({c}) we plot the frequency dependence of the circuit as a function of length $\ell$ for a width of $w=12.5$ $\mu$m.  Similarly, in Fig.~\ref{fig:Frequency_Dependence}({d}) we show the frequency dependence as a function of width for a length of $\ell =1.5$ $\mu$m.
The experimental results (grey circles) are in excellent agreement with GPE simulations~\cite{refSupp,Dennis:2012} (orange diamonds) in both cases.

\textit{Acoustic circuit model.---} Since a BEC of ultracold atoms forms a compressible superfluid, the macroscopic parameters of the system (e.g. circuit geometry, average fluid density and pressure) are amenable to  modeling using the theory of acoustics~\cite{Morse1986}. This methodology underpins the modeling of a variety of compressible fluid phenomena, with diverse applications to cosmology~\cite{Percival:2010}, plasma physics~\cite{Rao:1990}, oceanography~\cite{Medwin:1998}, seismology~\cite{Artru:2004}, and architectural design~\cite{Mehta:1999}.

\begin{figure}[t!]
\includegraphics[width=\linewidth, trim = 0 0 0 0, clip ]{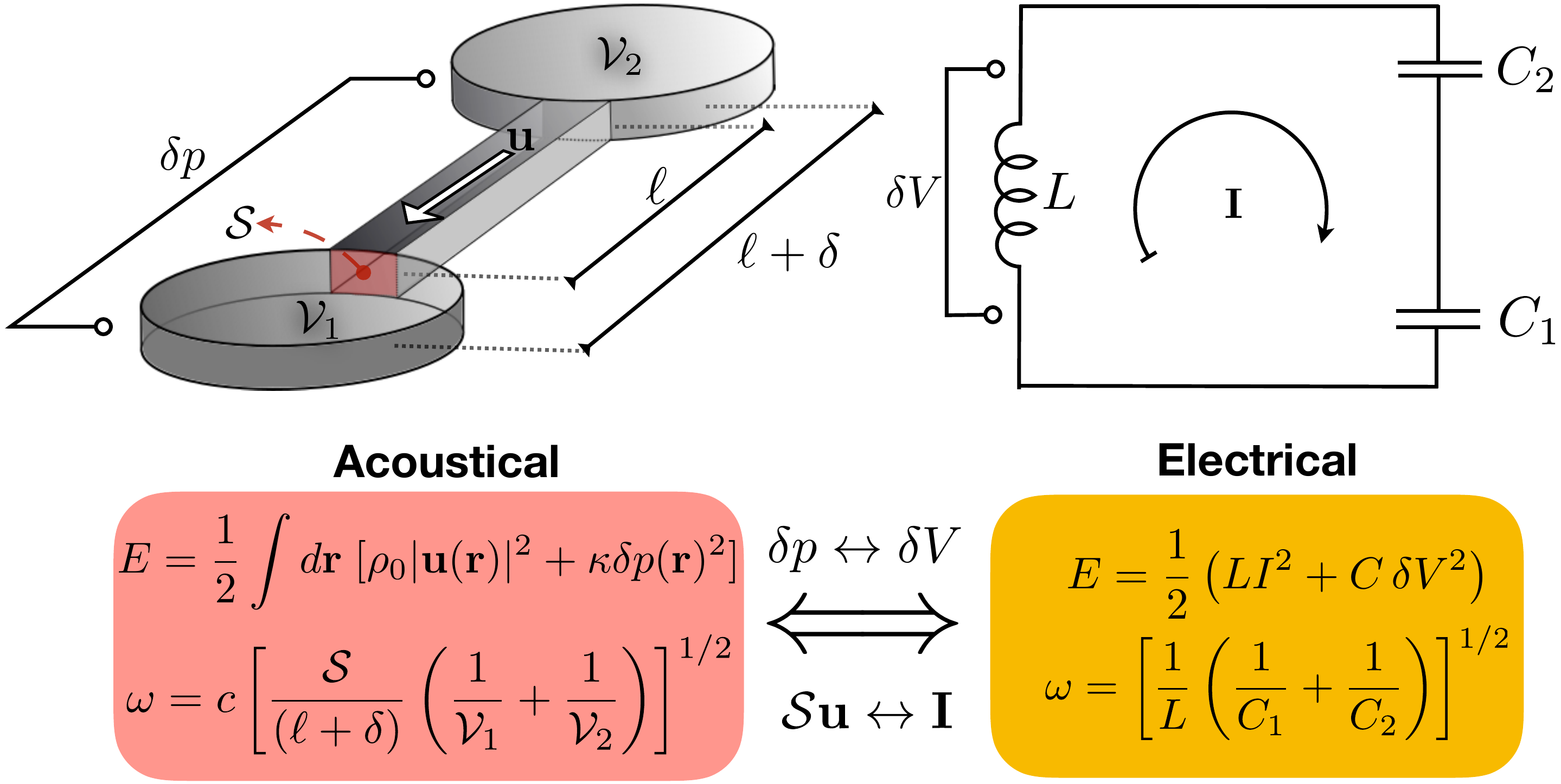}%
\caption{ {Acoustic Helmholtz oscillator and equivalent $LC$ circuit.} (Left) A Helmholtz superfluid oscillator consists of two reservoirs of volume ${\cal V}_1$ and ${\cal V}_2$ coupled by a channel of cross-sectional area $\mathcal{S}$ and length $\ell$. A pressure differential $\delta p$ leads to fluid exchange at velocity $\mathbf{u}$ between the reservoirs. The exchange of energy $E$ between the kinetic and potential terms leads to oscillations at frequency $\omega$, dependent on $\mathcal{S},~\mathcal{V}$, and the effective channel length $(\ell + \delta)$, with the end correction $\delta$ accounting for fluid flow inside the reservoirs. (Right) Mapping $\delta p$ to voltage $\delta V$ and volume velocity $\mathcal{S}\mathbf{u}$ to current $\mathbf{I}$ results in an equivalent $LC$ circuit model for the system.  
}
\label{fig:HelmholtzCircuit}
\end{figure}

A lumped-elements acoustic model can be realized using the correspondence between the kinetic and potential energy stored within the sound waves in a superfluid, and the electrical energy in an $LC$ circuit, as illustrated in Fig.~\ref{fig:HelmholtzCircuit}.  Specifically
 \begin{equation}
E = \tfrac{1}{2} (L I^2 + CV^2 ) =  \tfrac{1}{2}\int \, d\mathbf{r} \left[\rho_0 |\mathbf{u}(\mathbf{r})|^2 + \kappa\, \delta p(\mathbf{r})^2 \right]
     \label{eqn:Atomtronics}
 \end{equation}
where $\rho_0$ is the superfluid mass density at hydrostatic equilibrium, $\textbf{u}(\mathbf{r})$ is the superfluid velocity field, $\delta p(\mathbf{r})$ is the pressure difference from hydrostatic equilibrium, and $\kappa = 1/(\rho_0 c^2)$ is the compressibility, which depends upon the speed of sound, $c$.  
Provided the wavelength of the sound is much larger than the characteristic length scales of the atomtronic device, all acoustic variables can be treated as constant over the dimensions of the device, leading to a lumped-elements description~\cite{Morse1986}.

Applying this methodology, it can be shown that the resonator frequency is given by~\cite{refSupp}
\begin{equation}
\omega = \frac{1}{\sqrt{L C}} = c\left[\frac{\mathcal{S}}{\ell + \delta} \left(\frac{1}{\mathcal{V}_1} + \frac{1}{\mathcal{V}_2}\right) \right]^{1/2}, \label{eq:acoustic_freq}
\end{equation}
where $L = \rho_0(\ell+\delta)/\mathcal{S}$ and $C = \kappa(1/\mathcal{V}_1 + 1/\mathcal{V}_2)^{-1}$. Here $\mathcal{V}_1$ and $\mathcal{V}_2$ are the volumes of the reservoirs, $\mathcal{S}$ is the cross sectional area of the channel connecting the two reservoirs, $\ell$ is the length of the channel, and $\delta \propto \sqrt{\mathcal{S}}$ is the so-called end correction due to the transport extending into the reservoirs~\cite{Morse1986,Ingard1953}. Notably, this frequency depends only on the geometry of the resonator, the speed of sound, and the end correction. 

To test this model, we determine the speed of sound by calculating the density-weighted average from the groundstate of the GPE~\cite{refSupp,Chiofalo:2000}, and determine the end correction by fitting Eq.~(\ref{eq:acoustic_freq}) to the oscillation frequencies predicted by the GPE simulations.  The frequency prediction made by the acoustic model is shown as the black dashed line in Fig.~\ref{fig:Frequency_Dependence}(c,d) and is in excellent agreement with the experiment.  Quantitatively, the residuals associated with the end correction fits to the GPE frequency vs.~length and frequency vs.~width data are $< 1.9\%$ and $<5.2\%$, respectively.  The agreement is even better for longer channels~\cite{refSupp}. These results demonstrate that once the end correction has been determined, the acoustic model can quantitatively model the superfluid circuit and detailed GPE modelling is not required. We find that the hydrodynamic treatment of the oscillator is essential to predicting the correct frequency; a direct application of an ordinary $LC$ circuit ~\cite{lee_analogs_2013,eckel_contact_2016}, without the end correction, leads to overestimates of the resonator frequency of up to a factor of 3~\cite{refSupp}.

\begin{figure}[t!]
\includegraphics[width=\linewidth, trim = 0 0 0 0, clip ]{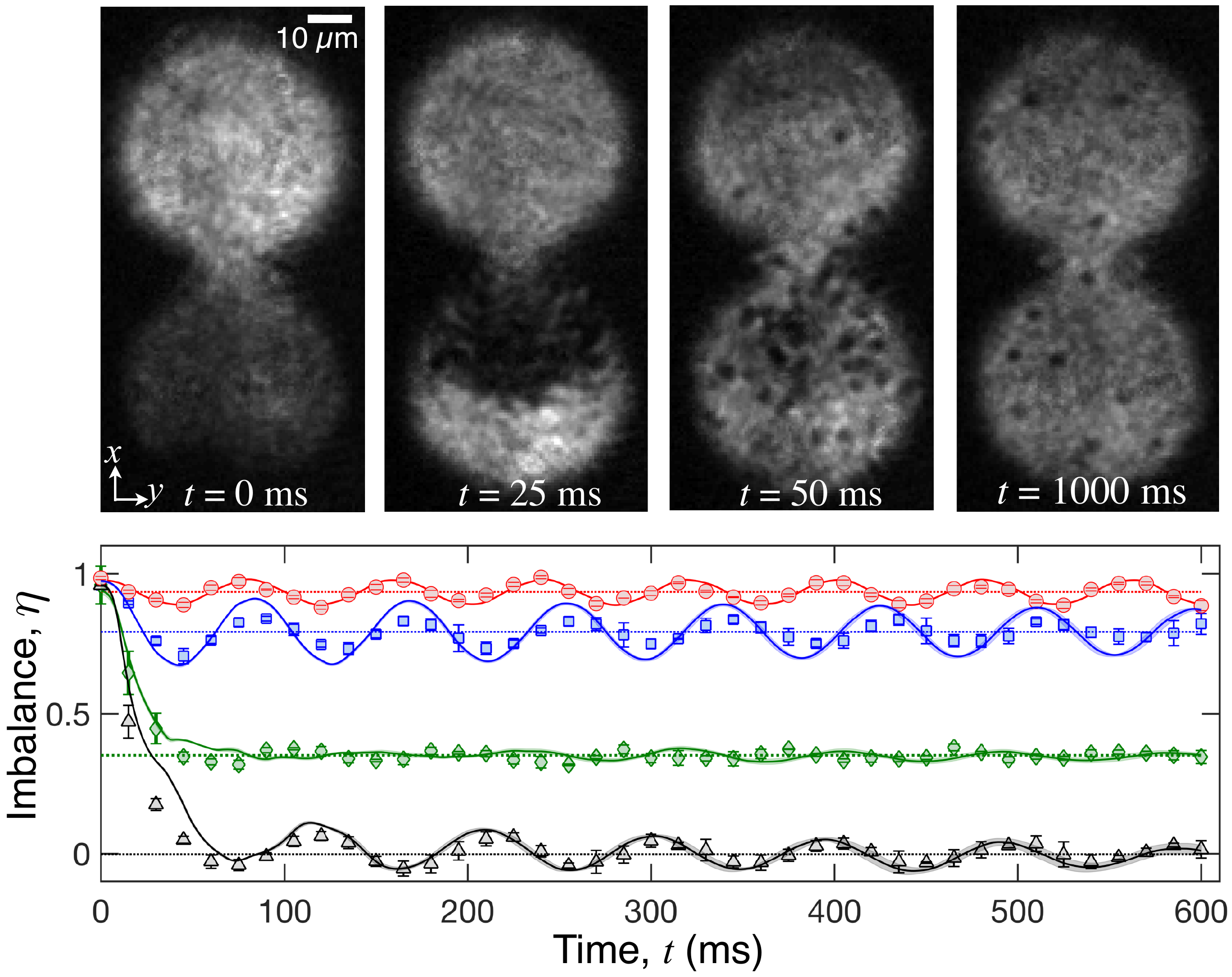}%
\caption{{Superfluid atomtronic oscillation regimes, for channel width $w=12.5~\mu$m, and channel length $\ell=1.5~\mu$m.} (Top row) 
Superfluid density dynamics, beginning from an initial $\eta=0.96(1)$ population imbalance. 
The Faraday images were taken following 3~ms time-of-flight expansion. In this regime, the large initial imbalance leads to phase-slip-induced dissipation; vortices can be clearly seen as dark holes in the density. (Bottom panel) Dynamics of the population imbalance for different initial imbalances. Red circles: $\eta = 0.03(1)$; blue squares: $\eta = 0.18(1)$; green diamonds: $\eta=0.60(5)$; black triangles: $\eta=0.96(1)$; continuous lines: GPE simulations. The data are offset for clarity, with horizontal lines denoting $\eta=0$. Uncertainties are 95\% confidence intervals estimated from five datasets.}
\label{fig:Experimental_Setup}
\end{figure}

\textit{Dissipative regime.---} We now turn to larger initial imbalances where we characterize the resistive damping of the superfluid flow~\cite{brantut_conduction_2012, burchianti2018connecting,krinner2013superfluidity,jendrzejewski2014resistive,spagnolli_crossing_2017,krinner2017two}. In contrast to a classical acoustic circuit, where resistivity originates from the effects of viscosity~\cite{Morse1986}, in a superfluid the resistance is expected to originate from nonlinear excitations such as vortex lines~\cite{eckel_contact_2016}, vortex rings~\cite{burchianti2018connecting,xhani2019critical}, or solitons~\cite{Meyer2017}, depending on the circuit geometry. Typical results for the dynamics of the population imbalance are shown in Fig~\ref{fig:Experimental_Setup}; above a critical imbalance $\eta_c\sim5\%$ we see clear evidence of vortices in the superfluid, and decay of the imbalance with time. At very large biases, the decay is a highly turbulent process in which a large density shock decays into a disordered distribution of vortices [Fig.~\ref{fig:Experimental_Setup}(a)]. After the dissipative phase, we find the system exhibits reproducible undamped oscillations~[Fig.~\ref{fig:Experimental_Setup}(b)] at long times~\footnote{The oscillation frequencies in Fig.~\ref{fig:Experimental_Setup} decrease with increasing initial bias. This is due to a reduced atom number ($\omega \propto N^{1/3}$~\cite{refSupp}); at sufficiently high biases the chemical potential is comparable to the DMD potential depth, meaning fewer atoms can be contained in the trapping potential.}.

We find that, after the initial resistive decay, the amplitude of oscillations is sensitive to the initial bias. In both the experimental and GPE data, the 60\% bias case shows almost complete suppression of oscillations, while the 96\% bias case exhibits oscillations $\sim5-10$\% in amplitude. In individual GPE simulations, the oscillation amplitude was found to be strongly sensitive to  the initial imbalance, atom number, and channel dimensions, displaying  no apparent trend against these parameters~\footnote{In the GPE simulations, altering the initial bias by as little as 5\% could lead to full recovery of the oscillations. Furthermore, the initial phase of the oscillation depended strongly on atom number in this regime, contributing to the reduced average amplitude, although this behavior was not apparent in the experimental data.}.

The behavior of the decay and subsequent reproducible oscillations seen in Fig.~\ref{fig:Experimental_Setup} motivates the inclusion of a phenomenological resistive element, $R$, in the acoustic circuit. Previous studies~\cite{Burchianti:2017,eckel_contact_2016} have presented evidence in favor of phase-slip models of dissipation~\cite{Avenel1997,jendrzejewski2014resistive,burchianti2018connecting}. This results in an acoustic conductance~\cite{refSupp}
\begin{equation}
G = {\cal A}\frac{n_{\textrm{1D}}\xi}{h \bar{n}^2},
\label{eqn:PhaseSlipG}
\end{equation} 
where ${\cal A}$ is a dimensionless scaling constant of order unity, $n_\textrm{1D}$ is the 1D density, and $\xi$ is the healing length measured $1~\mu$m before the channel exit.

\begin{figure}[t]
\includegraphics[width=\linewidth, trim = 0 0 0 0, clip]{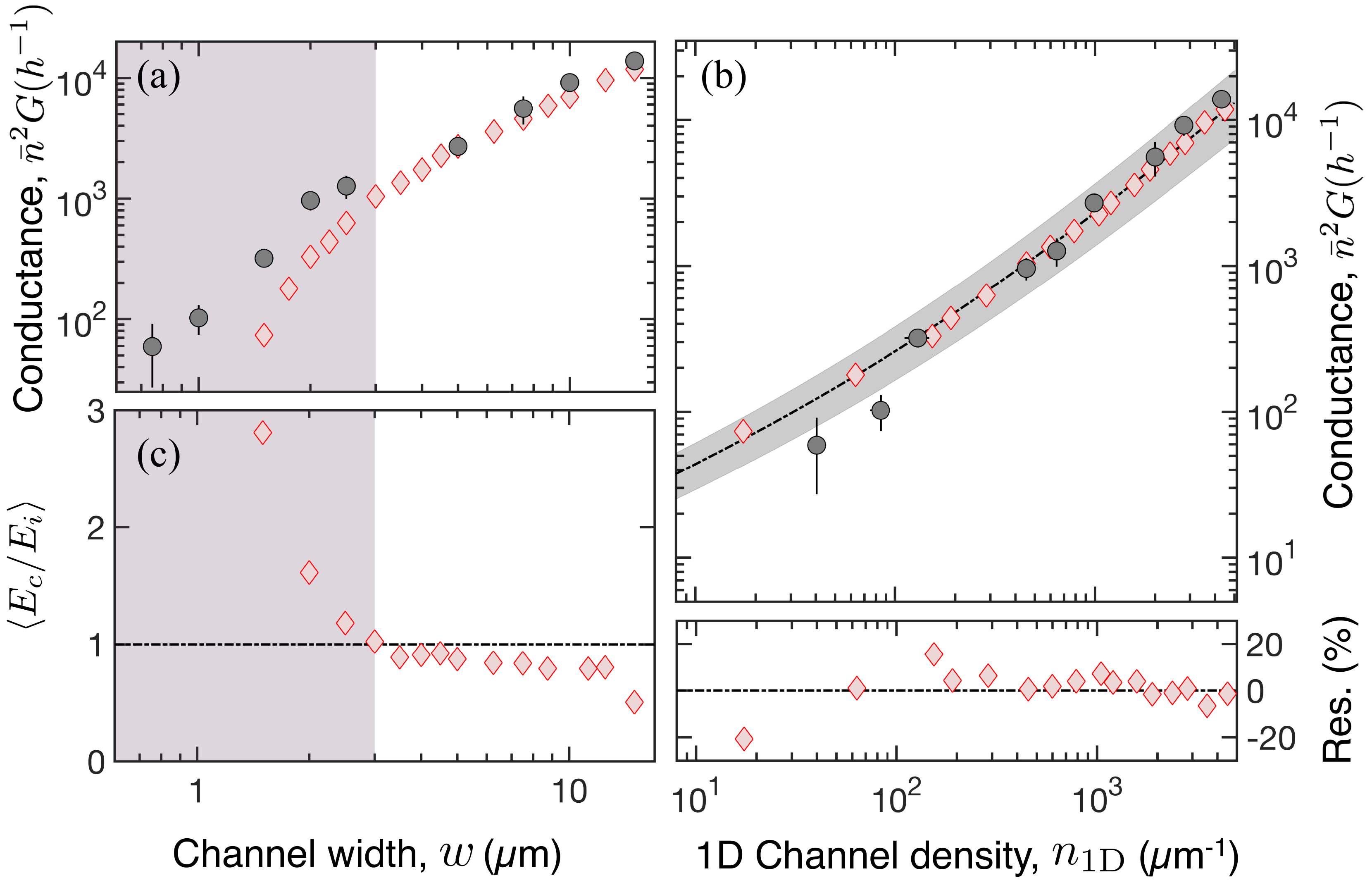}%
\caption{{(a)} Channel conductance as a function of programmed channel width, upon which $n_{\textrm{1D}}$  depends, for the GPE simulations (red diamonds) and experimental data (grey circles), for initial bias $\eta \sim 1$ and fixed channel length $\ell = 10\mu$m. The purple shaded area serves to guide the eye, highlighting a change in the conduction power law for small channel widths. For channel widths $w<3\mu$m, the resolution of the DMD projection results in the channel floor rising as the width decreases, further reducing the density in the channel~\cite{refSupp}. {(b)} Channel conductance as a function of the 1D channel density $n_\textrm{1D}$; the data fall on the same trend line. The power law fit (black dot-dashed line) demonstrates a deviation from the phase-slip model (see text). The light-grey region indicates 95\% confidence intervals for the fit, and residuals are shown below. {(c)} Ratio of compressible and incompressible energies averaged over the resistive dissipation regime (exponential decay regime in Fig.~\ref{fig:Experimental_Setup}(b)), showing that compressible excitations are dominant for small channels.}
\label{fig:Resistive_Decay}
\end{figure}

We have performed experiments and GPE simulations to measure the conductance $G = 1/R$ as a function of channel width $w$ for a fixed length of $\ell=10$~$\mu$m and an initial imbalance of $\eta\sim 1$~\cite{refSupp}. Our experiments probe channel widths over the range $0.75~\mu\textrm{m} - 15~\mu$m~\footnote{We experimentally observe small but finite transport between the reservoirs even in the absence of a channel due to condensate distillation~\cite{shin2004distillation}.}, resulting in the one-dimensional channel densities $n_\textrm{1D}$ varying over nearly two orders of magnitude, allowing a stringent test of the phase-slip model in high-bias conditions. We determine the conductance by fitting the initial decay of the imbalance $\eta(t)$ as seen in Fig.~\ref{fig:Experimental_Setup}(b) to an $RC$ decay model~\cite{refSupp}, and the results are shown in Fig.~\ref{fig:Resistive_Decay}.  For ease of comparison with Refs.~\cite{eckel_contact_2016,jendrzejewski2014resistive, burchianti2018connecting}, the \emph{chemical conductance} $G_{\mu}$ is plotted vs programmed DMD channel width and $n_{\textrm{1D}}$ in Figs.~\ref{fig:Resistive_Decay}({a}) and \ref{fig:Resistive_Decay}({b}), respectively~\footnote{The effective width is smaller than the programmed width for small channels due to the finite projection resolution~\cite{refSupp}.}. The chemical conductance is related to the acoustic conductance through $G_{\mu} = \bar{n}^2G$~\cite{refSupp}. The experimental results (grey circles) are in good agreement with the GPE simulations (red diamonds). 

To test whether the phase-slip model accurately predicts the conductance, we fit the GPE conductance data with Eq.~(\ref{eqn:PhaseSlipG}) by assuming a power-law dependence of the conductance on the 1D density, $G \propto {\cal A} (n_\textrm{1D}\xi)^{\alpha}$, where the model predicts $\alpha = 1$. The resulting fit and fit residuals are shown with a dot-dashed curve in Fig.~\ref{fig:Resistive_Decay}({b}). Fitting the conductance vs.~$n_\textrm{1D}\xi$, where $\xi$ is determined from the GPE simulations, results in $\alpha = 1.18(3)$ and ${\cal A} = 1.8(4)$, inconsistent with the phase-slip model prediction. Fitting vs.~$n_\textrm{1D}$ only, as was done in previous work~\cite{eckel_contact_2016,burchianti2018connecting}, gives a similar disagreement with $\alpha = 1.11(4)$. A fit to the experimental data, using the GPE-derived healing lengths, results in $\alpha = 1.2(1)$ and ${\cal A} = 1.2(11)$, consistent with the GPE results at the 95\% confidence level. 

We observe a qualitative difference in the dynamics for small and large channels~\cite{refSupp}. Small channel widths, $w \lesssim 8\xi \sim 4\mu$m, cannot support vortex dipoles; the topological excitations are instead solitonic vortices or solitons~\cite{brand2002solitonic}, which can rapidly decay to compressible (sound) excitations. Comparison of the relative magnitude of compressible excitations (sound) and incompressible excitations (vortices) for the varying channel widths, as determined from GPE simulations~\cite{refSupp}, is shown in Fig.~\ref{fig:Resistive_Decay}(c). Dissipation is sound dominated for small channels, whereas for larger channels sound and vortices contribute roughly equally. Despite these differences, we find a similar power law when fitting only the channels $w \leq 4\mu$m [$\alpha = 1.3(2)$], or larger channels $w > 4\mu$m [$\alpha = 1.19(5)$], indicating the conductance is independent of the nature of the excitations.

The consistent deviation from the phase-slip model suggests that understanding the more complex role of compressible excitations and excitation interactions will be required to fully describe the high-bias resistive decay of atomtronic circuits. We note that Ref.~\cite{burchianti2018connecting} observed a decreasing conductance with increasing bias, consistent with interactions between excitations, as might be anticipated for a high shedding rate. Indeed, for a superfluid the drag force is expected to be nonlinear for sufficiently high flow velocities~\cite{winiecki2000vortex}, or sufficiently wide obstacles, as is the case in classical hydrodynamics~\cite{frisch1995turbulence}. Establishing empirical laws for these fluid aspects, as has been done for classical fluids~\cite{Morse1986}, will thus be important for completely understanding the resistive behavior of superfluid circuits.

\textit{Outlook.---} Our comprehensive study of transport between superfluid reservoirs connected by a linear channel provides a rigorous foundation for a lumped-element acoustic approach to atomtronic circuits. In the nondissipative regime, the frequency and amplitude of oscillations for any channel geometry and moderate initial bias can be completely determined by the geometry, speed of sound, and a single fitted parameter, the end correction, which appears similar to that of a classical fluid. 

Although the dissipative regime fails to conform to a simple phase-slip picture, our results indicate that $n_\textrm{1D}$ can still be used to estimate the conductance, as was done previously~\cite{burchianti2018connecting,eckel_contact_2016,jendrzejewski2014resistive}. A more detailed understanding of the resistive mechanism requires a full characterization of the underlying turbulent dynamics. 

The intermediate bias case shown in Fig.~\ref{fig:Experimental_Setup}(b) also highlights the need for a more comprehensive study of conductance vs. initial bias; although the reduced oscillation amplitude at intermediate bias could be reproduced with previously developed shunted Josephson junction resistor models~\cite{burchianti2018connecting,eckel_contact_2016}, this requires the critical current to vary nonmonotonically (perhaps randomly) with initial bias. We found the behavior could not be explained by the resonant excitation of a higher-order acoustic mode. Additionally, analysis of the incompressible and compressible decomposition showed no significant discrepancy with the 100\% bias case, suggesting this is not due to enhanced conversion of the initial bias energy into vortex interaction energy. In the spirit of a circuit model, this behavior might be explained by a random relative phase acquired between the reservoirs during the resistive decay phase. Alternatively, the behavior might originate from the interaction between the fundamental resonator mode and the turbulent background, and require theories for the non linear ring down of acoustic resonators.

Our work is a step towards more complex atomtronic circuits, and can be applied to the design of guided superfluid inertial sensors of superfluid helium~\cite{Avenel1997,Hoskinson2006} and exciton-polariton condensates~\cite{Sanvitto2016}.  Future work will include the realization of other passive atomtronic circuits, such as simple high- or low-pass filters, which have been developed for other acoustic systems~\cite{Morse1986}. Such elements may be useful for modulating the coupling between superfluids and mechanical elements, such as superfluid helium hybrid systems~\cite{harris2016laser}. Our approach provides a framework for the future combination of passive elements with active ones such as atomtronic transistors~\cite{caliga2016transport}.

\begin{acknowledgments}
We thank A.~S.~Bradley, J.~D.~Close, and A.~C.~White for useful discussions. This research was funded by the Australian Research Council Centre of Excellence for Engineered Quantum Systems (Project No. CE1101013 and No. CE170100009), and Australian Research Council Discovery Projects No.~DP160102085. This research was also partially supported by the Australian Research Council Centre of Excellence in Future Low-Energy Electronics Technologies (Project No. CE170100039) and funded by the Australian Government. T.A.B. and G.G. acknowledge the support of an Australian Government Research and Training Program Scholarship. S.S.S. received funding from Australian Research Council Discovery Projects No.~DP160104965 and No.~DP150100356.
\end{acknowledgments}

%

\clearpage
\newpage

\section{Supplemental Material}

\setcounter{equation}{0}
\setcounter{figure}{0}
\setcounter{table}{0}
\setcounter{page}{1}
\makeatletter
\renewcommand{\theequation}{S\arabic{equation}}
\renewcommand{\thefigure}{S\arabic{figure}}

\subsection*{Atom trapping setup}
\begin{figure}[t]
\center \includegraphics[width=\linewidth]{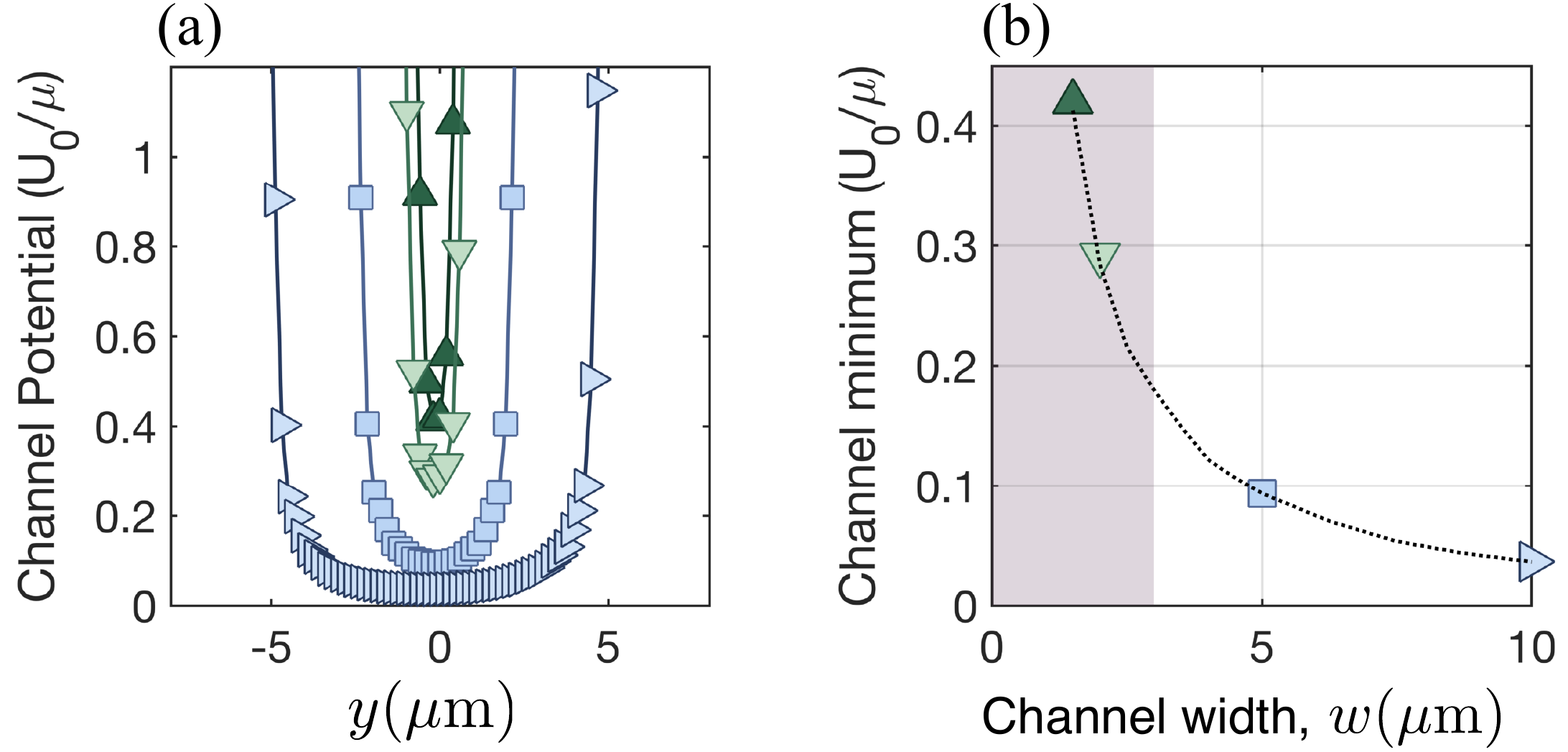}
\caption{{Variation of the channel potential minimum.} As described in the text, the floor of the potential rises for the smallest channels due to the projection resolution of the DMD. (a) Cross sections of the total potential $V(0,y,0)$ as a fraction of the chemical potential $\mu \sim 76$~nK for a number of programmed channel widths: $10$~$\mu$m (blue right-facing triangles),  $5$~$\mu$m (blue squares),  $2.5$~$\mu$m (green downwards-facing triangles), and  $1.5$~$\mu$m (dark green upwards-facing triangles). (b) For small channel widths the potential minimum rises significantly; the purple shaded region indicates where the data deviates from a linear trend, as shown in Fig.~4 of the main text.}
\label{fig:ChannelMinimum}
\end{figure}

The experimental setup, previously described in Ref.~\cite{gauthier_direct_2016}, produces $^{87}$Rb Bose-Einstein condensates (BECs) in the $|F=1,m_{F}=-1\rangle$ state. The atoms are confined in a plane by a red-detuned optical sheet beam of 1064 nm light, with radial and vertical harmonic trapping frequencies $(\omega_r,\omega_z)/2\pi =(6,300)$~Hz. The atoms are supported against gravity by a vertical magnetic field gradient of 30.5\,G/cm, resulting in a planar superfluid. The levitation field is produced by an unbalanced quadrupole magnetic field, which additionally results in a 80\,G field in the vertical direction. Configurable hard-walled confinement in the horizontal plane is provided by direct projection of 532\,nm blue-detuned light, patterned using a Digital Micromirror Device (DMD) that can project patterns with a resolution of $\sim 650$\,nm full-width-half-maximum~\cite{gauthier_direct_2016}. For the smallest channels, the finite projection resolution results in some light within the channel, raising the channel potential minima; see Fig.~\ref{fig:ChannelMinimum}. \emph{In situ} images of the BEC in the combined potential are shown in Fig.~1 of the main text.

\subsection*{Imaging and measurement of reservoir populations}
We determine the number of atoms with dark-ground Faraday imaging, using light detuned by 220\,MHz from the $^{87}$Rb $|F=1\rangle\rightarrow|F'=2\rangle$ transition in the presence of the 80\,G residual vertical magnetic field. The imaging system has a $52.6\times$ magnification and results in images with a measured resolution of 960(80)\,nm full-width-at-half-maximum at 780\,nm illumination~\cite{gauthier_direct_2016}. The number of atoms in each reservoir, denoted $N_1$ and $N_2$, is determined by counting the number of atoms on each side of the midpoint of the channel.

\subsection*{Biasing the reservoir populations}
To generate a non-zero imbalance between the reservoir populations, $\eta = \left(N_{1}-N_{2})/(N_{1}+N_{2}\right)$, during the evaporation sequence we apply an additional linear magnetic gradient along the long  ($x$) axis of the trap. The resulting magnetic potential increases the number of atoms in one reservoir at the expense of the other, resulting in a chemical potential difference between the reservoirs $\Delta \mu$ when the biasing potential is removed. Due to the limited depth of the DMD potential, the atom number varies with the applied bias.  The total atom number is $N\sim~2\times 10^6$ for the smallest imbalances and $N\sim~7.5\times 10^5$ for the $\eta \sim 1$ initial bias case (larger biases result in reduced atom number with higher filling). The bias is suddenly turned off at time $t=0$\,ms. After some evolution time, a destructive, dark-ground Faraday image is taken to measure the reservoir populations.

\subsection*{Superfluid acoustic theory} 
A trapped BEC can be modelled by the Gross-Pitaevskii equation (GPE):

\begin{equation}
i\hbar\frac{\partial \psi(\mathbf{r},t)}{\partial t} = \left[-\frac{\hbar^2}{2m}\mathbf{\nabla}^2 + V_{\rm ext}(\mathbf{r},t) + g|\psi(\mathbf{r},t)|^2\right]\psi(\mathbf{r},t),
\label{eqn:GPE}
\end{equation}
where $V_{\rm ext}(\mathbf{r},t)$ is the external confining potential and $g = 4\pi \hbar^2a_s/m$ is the interatomic interaction strength for $s$-wave scattering length $a_s$ and atomic mass $m$. The BEC wave function $\psi(\textbf{r},t)$ is normalized to the total number of particles, $N$. Adopting the Madelung transformation $\psi(\mathbf{r},t) = \sqrt{n(\textbf{r},t)} e^{i\theta(\mathbf{r},t)}$, the GPE can be reformulated into a hydrodynamic-like description for the velocity field $\mathbf{u}(\mathbf{r},t)= \hbar\mathbf{\nabla}\theta(\mathbf{r},t)/m$ and mass density $\rho(\mathbf{r},t) = m \, n(\mathbf{r},t)$:
\begin{gather}
    \partial_t \rho + \nabla \cdot (\rho \mathbf{u}) = 0, \label{eqn:continuity}\\
        \rho (\partial_t \mathbf{u} + (\mathbf{u} \cdot \nabla) \mathbf{u}) = -\nabla p + \nabla \cdot \boldsymbol \Sigma + \mathbf{f}, \label{eqn:quantumEuler}
\end{gather}
where $p(\rho) = g\rho^2/2 m^2$ is the pressure field, $\mathbf{f} = -\rho \nabla V_{\rm ext} /m$ defines external forces, and $\boldsymbol \Sigma = (\hbar/2 m)^2 \nabla \otimes \nabla \ln \rho$ is the quantum stress tensor. Note that the quantum Euler equation, \eqnreft{eqn:quantumEuler}, assumes no vortices are present. In the presence of vortices it will contain additional terms.    Assuming small perturbations about a uniform hydrostatic equilibrium $V_{\rm ext} = 0$, $\rho = \rho_0 + \delta\rho(\mathbf{r},t)$, $p = p_0 + \delta p(\mathbf{r},t)$, and $\mathbf{u} = \delta\mathbf{u}(\mathbf{r},t)$, \eqnsreft{eqn:continuity}{eqn:quantumEuler} yield a wave equation for dispersive density waves:
\begin{equation}
    \partial_t^2 \delta \rho = (\rho_0 g/m^2) \nabla^2 \delta\rho  - (\hbar/2m)^2 \nabla^4 \delta\rho,
\end{equation}
where higher-order terms in $\delta\rho$ have been neglected. For excitations with wavelength $\lambda$, the second term on the RHS can be neglected for $\lambda \gg \xi /2$, where $\xi = \hbar/\sqrt{\rho_0 g}$ is the superfluid healing length. The resulting description yields ordinary (nondispersive) sound waves travelling at speed $c = \sqrt{ \rho_0 g/m^2}$. Therefore, for small amplitude and long wavelength perturbations, the BEC behaves as an ideal compressible fluid, which is barotropic ($p = p(\rho)$ only). In a barotropic fluid, the energy of a sound wave is given by~\cite{Morse1986}
\begin{equation}
E = \frac{1}{2} \int d\mathbf{r} \, \left[\rho_0 |\mathbf{u}(\mathbf{r})|^2 + \kappa\, \delta p (\mathbf{r} ) ^2\right],
\label{eqn:acoustic_energy}
\end{equation}
where $\kappa = 1/\rho_0 c^2$ is the compressibility.

\subsection*{Superfluid acoustic circuit}  
In the $LC$ circuit analogy, the channel behaves as a kinetic inductor, and each reservoir behaves as a capacitor (see Fig.~2 of the main text).
The volume velocity $\mathcal{S}\mathbf{u}$ assumes the role of the current $\mathbf{I}$, and changes in pressure $\delta p$ assume the role of the potential difference $\delta V$. 
We assume that the channel contains negligible potential energy, and the reservoirs contain negligible kinetic energy. If the wavelength of the sound is larger than all the characteristic scales of the device, i.e., $\lambda \gg \{\ell,\mathcal{S}^{1/2},\mathcal{V}^{1/3}\}$, $\mathbf{u}$ and $\delta p$ may be considered uniform over the relevant volumes. Integrating the right hand side of Eq.~(\ref{eqn:acoustic_energy}) yields
 \begin{align}
 L &= \frac{\rho_0(\ell + \delta)}{\mathcal{S}},  & C_n & = \kappa \mathcal{V}_n.
 \end{align}
Here $(\ell +\delta)$ is the \emph{effective} channel length. The end correction $\delta$ accounts for the extension of the superfluid flow beyond the physical channel length $\ell$; in the electronic analogy, $\delta$ accounts for the residual inductance of the reservoirs. In general, $\delta \propto \sqrt{\mathcal{S}}$; for example, a circular channel area $\mathcal{S}$ typically yields $\delta \sim 0.96\sqrt{\mathcal{S}}$~\cite{Morse1986,Ingard1953}. Although $\delta$ can be calculated analytically for some simple geometries~\cite{Ingard1953}, in general it must be determined empirically or by detailed numerical modelling. A rigorous calculation of the end correction requires a solution to the problem of sound transmission through the resonator aperture, which is complicated by the fact that the velocity distribution in the aperture is not known \emph{a priori}~\cite{Ingard1953}.

For two capacitors in series, the total capacitance of the circuit is given by $1/C = 1/C_1 + 1/C_2$, and the resonator frequency Eq.~(2) of the main text follows from $\omega = 1/\sqrt{L C}$. Note that the above relationship between capacitance and volume can be equivalently obtained by considering the potential energy stored in the circuit due to a chemical potential difference (see below).

\subsection*{Comparison with an electronically-motivated model}
\begin{figure}[ht]
\center \includegraphics[width=\linewidth]{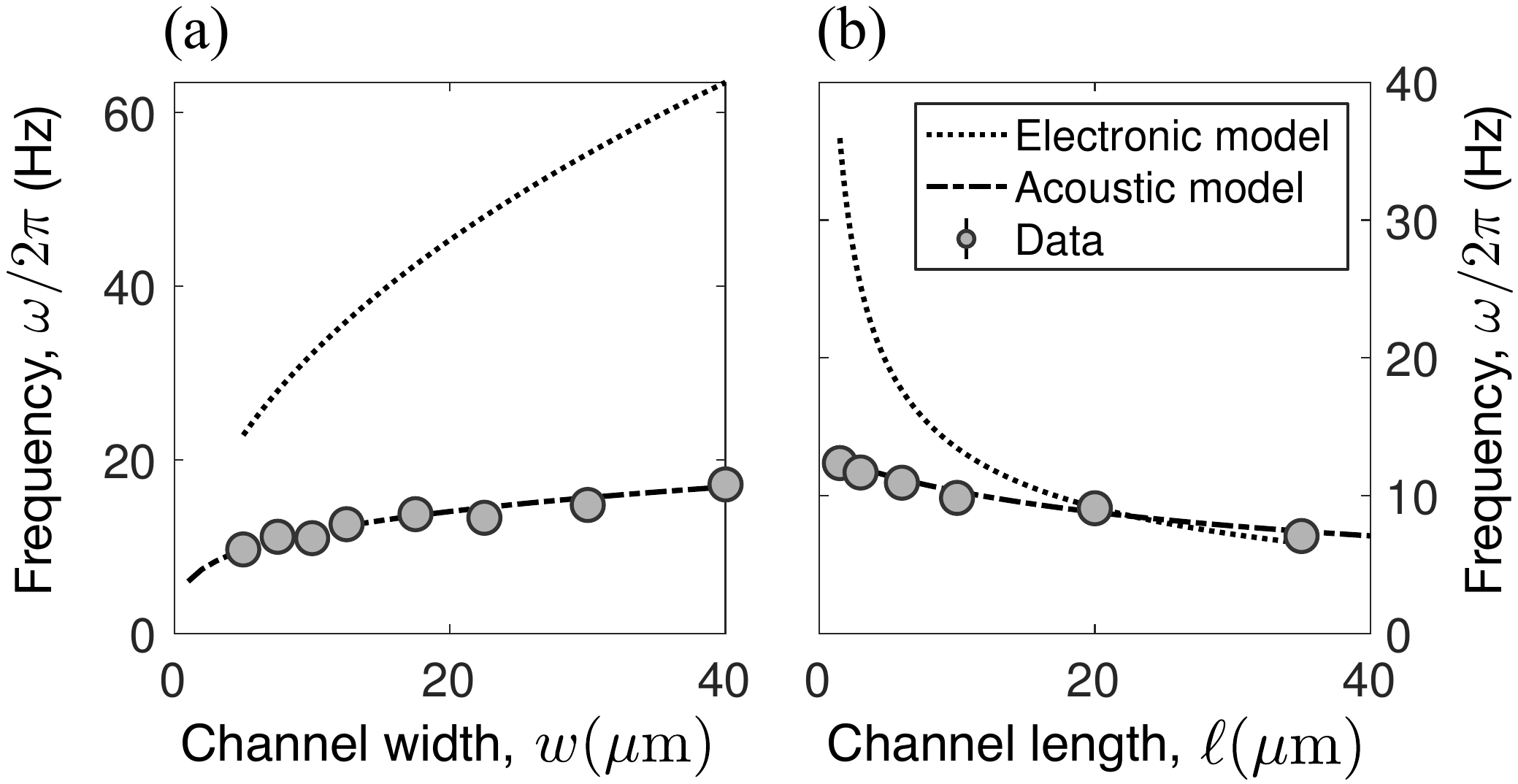}
\caption{(a) Comparison between the electronically-motivated circuit model of Refs.~\cite{lee_analogs_2013,eckel_contact_2016} and the acoustic model for fixed $\ell = 1.5$~$\mu$m channel length and varying width, as in Fig.~1(a) of the main text. The electronic model considers a channel inductance only (see text), which neglects the `contact' inductance due to fluid flow inside the reservoirs, leading to a poor prediction of the frequency. (b) Similar results are seen for a fixed channel width ($w = 10$~$\mu$m) and a varying channel length. Here the acoustic model gives quantitative agreement for all experimentally-probed lengths, whereas the electronic model only quantitatively predicts the correct frequency in the long-channel limit.}
    \label{fig:NIST_model compare}
\end{figure}

Reference~\cite{lee_analogs_2013} used Kirchoff's law to develop a simple model of the atomtronic circuit, where the kinetic inductance is determined solely from the properties of the channel, resulting in $L_E = 2m\ell/n_\textrm{1D}$. In Fig.~\ref{fig:NIST_model compare} this electronically-motivated model and the acoustic model are both compared with the experimental data. For the shortest length channels in Fig.~\ref{fig:NIST_model compare}(a) the electronic model overestimates the frequency by more than a factor of three for the $40$~$\mu$m-width channel. The electronic model is only suitable in the long-channel length regime where the channel's contribution to the inductance dominates, as can be seen in Fig.~\ref{fig:NIST_model compare}(b). These results support our conclusion that quantitative modelling of atomtronic circuits requires an acoustic approach.

\subsection*{Oscillator frequency dependence on atom number}
\begin{center}
\begin{figure} [ht]
\center \includegraphics[width=0.6\linewidth]{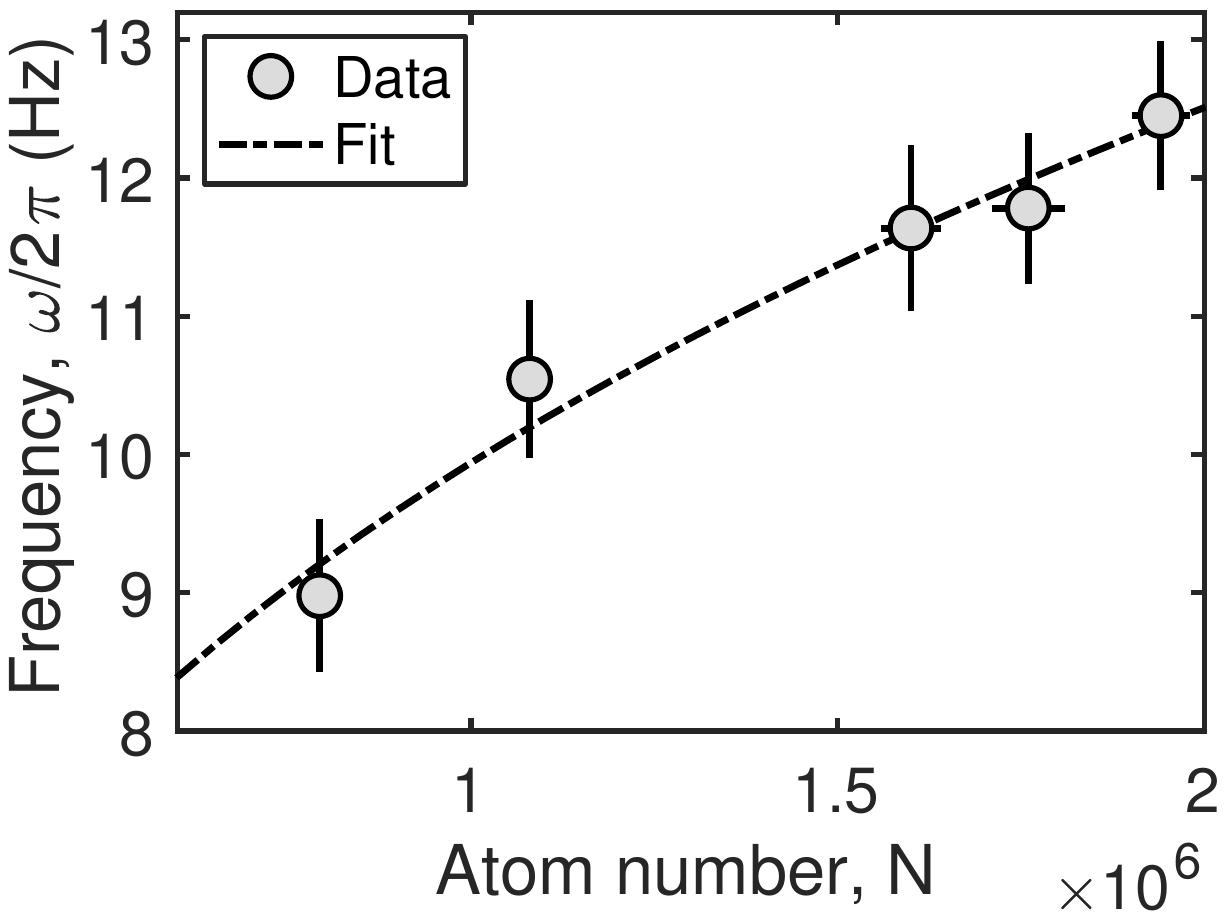}
\caption{Oscillator frequency dependence on atom number. The experimental data for $w = 12.5$~$\mu$m and $l=1.5$~$\mu$m is shown. The black dash-dot line is a power law fit to the data.}
\label{fig:NumberFrequencyDependence}
\end{figure}
\end{center}

In the nondissipative regime, the oscillation frequency for a given geometry (Eq.~(2) of the main text) varies with atom number through the speed of sound $c$. For harmonic trapping in the $z$-direction and uniform trapping in $x$ and $y$, the Thomas-Fermi approximation leads to $c\propto N^{1/3}$~\cite{gauthier2018negative}, and hence $\omega \propto N^{1/3}$. Due to the varying area of the trapping potential with the channel width and length, the initial atom number changes with the trap in the experiment. In order to consistently compare oscillation frequencies for the different channel widths and lengths shown in Fig.~1 of the main text, multiple experimental data sequences at each condition were taken, where the atom number was varied. The frequency dependence was then extracted through a power-law fit $\omega(N) = a N^{b} + \omega_0$. Figure~\ref{fig:NumberFrequencyDependence} shows examples of the experimental data and fit, for a trap geometry with $w = 12.5$~$\mu$m and $\ell=1.5$~$\mu$m, resulting in $a = 0.09(7)~$Hz, $b = 0.33(6)$, and $\omega_0 \simeq 0$. The frequencies plotted in Fig.~1 of the main text were determined by substituting $N=2\times10^6$ into these fitted functions; the error bars are 95\% confidence intervals from the fit.

\subsection*{Numerical simulation: nondissipative regime}
We directly modelled the experimental procedure using the 3D GPE, Eq.~(\ref{eqn:GPE}). The data shown in Fig.~1 of the main text was obtained via the following simulation procedure. Imaginary time propagation~\cite{Chiofalo:2000} was used to find the GPE groundstate for $N = 2 \times 10^6$ and potential $V(\textbf{r}) = V_\textrm{ext}(\textbf{r}) + V_\text{bias}(x)$.  Here $V_\textrm{ext}(\textbf{r})$ is the 3D potential formed from the red-detuned optical sheet and blue-detuned DMD-patterned light, accounting for the finite resolution of the projection system, the effect of gravity, and the counterbalancing levitation magnetic field, and $V_\text{bias}(x) = B x$. Values of $B$ were chosen between $2.5 \times 10^{-28}$~J/m and $2 \times 10^{-27}$~J/m, which resulted in initial conditions with a relative number imbalance between 0.5\% and 5\%. These initial conditions were evolved using Eq.~(\ref{eqn:GPE}) for $t = 0.6$\,s of simulation time with the bias turned off --- i.e.  $V(\textbf{r}) = V_\textrm{ext}(\textbf{r})$. Simulations were performed using the open-source software package XMDS2 \cite{Dennis:2012}, with an adaptive 4th--5th order Runge-Kutta interaction picture algorithm on a grid of $512\times 256\times16$ points with $\Delta x, \Delta y \sim 0.2$ $\mu$m and $\Delta z \sim 0.75$ $\mu$m. For a given trap geometry and initial bias, the oscillation frequency $\omega$ was determined by fitting a sinusoid $y = a \sin(\omega t + b)$ to the relative population difference $\eta(t) = (N_1(t) - N_2(t)) / (N_1(t) + N_2(t))$, where $N_1(t) = \int_{-\infty}^0 dx \int dy dz \, |\psi(\textbf{r})|^2$ and $N_2(t) = \int_{0}^\infty dx \int dy dz \, |\psi(\textbf{r})|^2$. 

For an atomic Josephson junction where the two reservoirs are separated by a tunneling barrier, the oscillation frequency $\omega$ depends on the initial bias $\eta_0$~\cite{albiez_direct_2005}. However, in the limit where the ratio of the tunneling energy to the on-site interaction energy tends to zero, as expected for a barrier-free channel, the sensitivity of the oscillation frequency to the initial bias vanishes. We investigated the dependence of the oscillation frequency on the bias by increasing the initial bias in the GPE simulations from $\eta_0 = 0$ to $\eta_0 = 0.05$; this was found to change $\omega$ by no more than 7\% (0.4\%) for the smallest (largest) channel widths, and 0.4\% (0.3\%) for the shortest (longest) channel lengths considered in Fig. 1. We account for this dependence by fitting a power law $\omega(\eta_0) = a \eta_0^b + \omega_0$ for each geometry. The frequency $\omega_0$ can be interpreted as the resonator frequency in the limit of infinitesimal initial bias. 

The GPE groundstate wave function, $\psi_0(\textbf{r})$, was used to estimate the average speed of sound in the circuit via $\bar{c} = \sqrt{g \bar{n}_0/ m}$, where the average number density $\bar{n}_0 = \bar{\rho}_0/m$ is obtained by averaging $|\psi_0(\textbf{r})|^2$ over the bulk of the condensate, defined as regions where $|\psi_0(\textbf{r})|^2 \geq 0.05 \max_\textbf{r} |\psi_0(\textbf{r})|^2$. The values of $\bar{c}$ obtained do not depend significantly on the precise region of averaging (averaging over $|\psi_0(\textbf{r})|^2 \geq 0.1 \max_\textbf{r} |\psi_0(\textbf{r})|^2$, the values differed from the quoted values by $\lesssim 4\%$). The values of $\bar{c}$  also did not vary substantially with geometry ($\simeq 15\%$ over the geometries considered). We therefore averaged these estimates of $\bar{c}$ across different channel widths (lengths) when comparing to the fixed channel length (width) data of Fig.~1 of the main text, and used this average as an input to the acoustic model.

\subsection{Acoustic phase-slip model}
In the main text, we fit the conductance data to the phase-slip model developed in previous studies~\cite{jendrzejewski2014resistive,burchianti2018connecting}. In the dissipative regime, this model predicts an Ohmic resistive current $I_R = \bar{n}^2G_{\textrm{PS}}\Delta \mu$, where $\Delta \mu$ is the chemical potential difference between the two ends of the channel. Assuming each resulting topological excitation contains $N_\textrm{ex}$ atoms, and estimating the rate of phase slips from the Josephson-Anderson relation $\gamma \simeq \Delta \mu/h$, $I_R \propto N_\textrm{ex}\gamma$. The conductance then resembles a classical resistor, linearly depending on $n_\textrm{1D}$ through $N_\textrm{ex} \simeq n_{\textrm{1D}}\xi$~\footnote{For a uniform system $n_{\textrm{1D}} = {\mathcal{S}}n_\textrm{3D}$.}, where $\xi$ is the healing length for the superfluid near the channel exit. This results in an (acoustic) conductance:

\begin{equation}
G = {\cal A}\frac{n_{\textrm{1D}}\xi}{h \bar{n}^2},
\label{eqn:PhaseSlipG}
\end{equation} 
where ${\cal A}$ is a dimensionless scaling constant of order unity. Comparison with the chemical conductance $G_{\mu}$ of Refs.~\cite{jendrzejewski2014resistive,burchianti2018connecting} can be made through $G_{\mu} = \bar{n}^2G$.

\subsection*{Numerical simulation: dissipative regime} The data shown in the bottom panel of Fig.~3 and Fig.~4 of the main text was generated using a similar simulation procedure to the nondissipative regime with the same simulation grid. Each solid line (shaded region) in Fig.~3 of the main text shows the average (standard deviation) of ten GPE simulations, with the initial atom number $N_i$ for each run stochastically chosen from a Gaussian distribution of mean $N$ and variance $(N/10)^2$, which accounts for the $\sim10\%$ shot-to-shot atom number fluctuations observed in the experiment. Due to the limited potential depth, the mean atom number varies with the applied bias; specifically, mean atom numbers (imbalances) used were $N = 2 \times 10^6$ ($\eta=0.03(1)$), $N = 1.6 \times 10^6$ ($\eta=0.18(1)$), $1.4 \times 10^6$ ($\eta=0.60(5)$), and $1.06 \times 10^6$ ($\eta=0.96(1)$). Both the $\eta=0.60(5)$ and $\eta=0.96(1)$ cases required a slight negative bias to attain a $
\eta=0$ equilibrium; without this, $\eta(t)$ oscillates around a nonzero offset due to an imbalance of vortices that are shed. 

For the data shown in Fig.~4 of the main text, a $\eta=1$ initial imbalance was prepared by finding the groundstate for a DMD potential that only included a single reservoir and $N=7.5 \times 10^5$ atoms to match the initial conditions of the experiment. Evolution under the full potential was simulated for times between $t=0.3$\,s and $t=24.5$\,s, dependent upon the choice of channel width (the channel length was fixed at $10$~$\mu$m). A small percentage of the atoms (no more than $6\%$) had sufficient energy to leave the potential during the simulation. These atoms travelled to the edge of the simulation grid, where they were removed with an absorbing boundary layer. The initial decay of the population imbalance is quantified via the decay constant $\tau$ and extracted by fitting the function $\eta(t) = (1-A) \exp(-t/\tau)+B \cos (\omega_0 t+E)+A$. This assumes that at higher biases, the undamped $LC$ oscillations are modified by an initial exponential decay associated with capacitative discharge. A similar fitting procedure was performed in Ref.~\cite{li_superfluid_2016}, which included functions that smoothly turned the capacitive discharge and $LC$ oscillation off and on, respectively; this was specifically of the form: $\eta(t) = [1 + \exp(\frac{t-t_c}{t_w})]^{-1} \exp(-t/\tau) + [1 + \exp(-\frac{t-t_c}{t_w})]^{-1}[B \cos(\omega_0 t+E)+A]$. We compared the $\tau$ obtained using both fitting functions and found broad agreement within statistical uncertainty. We therefore opted to use the former fit, as it contained fewer parameters. For the smallest channel width considered, 1.5~$\mu$m, a simple exponential fit was used to extract the decay constant since in this case the decay was quite slow.

\subsection*{Determining the conductance of the channel}
Following Ref.~\cite{eckel_contact_2016}, the  conductance can be calculated from the decay constant $\tau$ through $G_{\mu} = C_{\mu}/\tau$, where the chemical capacitance $C_\mu = \bar{n}^2 C \approx 3N/4\mu $ (see below). The acoustic conductance is simply $G_{\mu}/\bar{n}^2$. The chemical potential $\mu$ was found by calculating the Thomas-Fermi groundstate for the unbiased potential filled with $N = 7.5\times 10^5$ atoms.

We define the average 1D density within the bulk of the $10$~$\mu$m-length channel as $n_\textrm{\textrm{1D}}(t) = \frac{1}{8\mu\textrm{m}}\int_{-4\mu\text{m}}^{+4\mu\text{m}} n_\textrm{\textrm{1D}}(x,t)$, where $n_{\textrm{1D}}(x,t) = \int dy dz \left|\psi(\textbf{r},t)\right|^2$. By taking the time average of $n_\textrm{\textrm{1D}}(t)$ from  the first turning point of $\eta(t)$, which is just after the large initial decay due to capacitative discharge, we obtain the time-averaged 1D channel density, which forms the horizontal axis for the simulation data in Fig.~4 of the main text.

The phase-slip model for the conductance requires an estimate of the 2D healing length. Assuming the density is uniform across the width of the channel, the healing length is given by $\xi = \hbar / \sqrt{m n_\text{2D} g_\text{2D}}$. Assuming a Thomas-Fermi inverted parabolic profile of radius $l_z$ in the $z$-direction, $g_\text{2D} \approx 3 g/ (5 l_z)$~\cite{gauthier2018negative}. We take $n_\text{2D}$ as the 2D density (computed by integrating out the $z$ direction) averaged over the channel length ($x \in [-4,4]$~$\mu$m) and channel width (values of $y$ where the density was greater than 10\% of its peak value), and then time-averaged as discussed above.

\subsection*{Determining the conductance from the experimental data}
For the experimental data $n_\textrm{1D}$ was averaged over the channel length for a combination of five sequential time-step samples once $\eta$ reached its steady-state value. The decay constant $\tau$ was determined through a simple exponential decay function $\eta(t) = (1 -A)\exp(-t/\tau) +A$. For each channel width the capacitance values from the GPE simulations were used as described above to extract $G_{\mu} = C_{\mu}/\tau$.

Measurements of the experimental conductance, as shown in Fig.~4 of the main text, were limited to non-zero values due to a finite rate of BEC distillation between the reservoirs. This is present even in the absence of a channel, with transport facilitated by the presence of the thermal cloud~\cite{shin2004distillation}.

\subsection*{Estimate of critical imbalance $\eta_c$} Nondissipative superfluid oscillations occur up to an amplitude of $\eta_c$ before shedding excitations, where $\eta_c = (N_1 - N_2)_\textrm{c}/N$. Since the current relates to the number imbalance via $I = d(N_{1}-N_{2})/dt =  N d\eta/dt $, assuming maximum-amplitude sinusoidal oscillations of the form $\eta(t) = \eta_c\textrm{sin}(2 \pi \omega t)$ and differentiating results in $I = 2 \pi \omega N \eta_c \textrm{cos}(2 \pi \omega t)$. Rearranging this equation at $t=0$ gives $\eta_{c} = I_{c} / (2 \pi \omega N)$.

\begin{figure}[!t]
\centering
\includegraphics[width = 0.8\columnwidth]{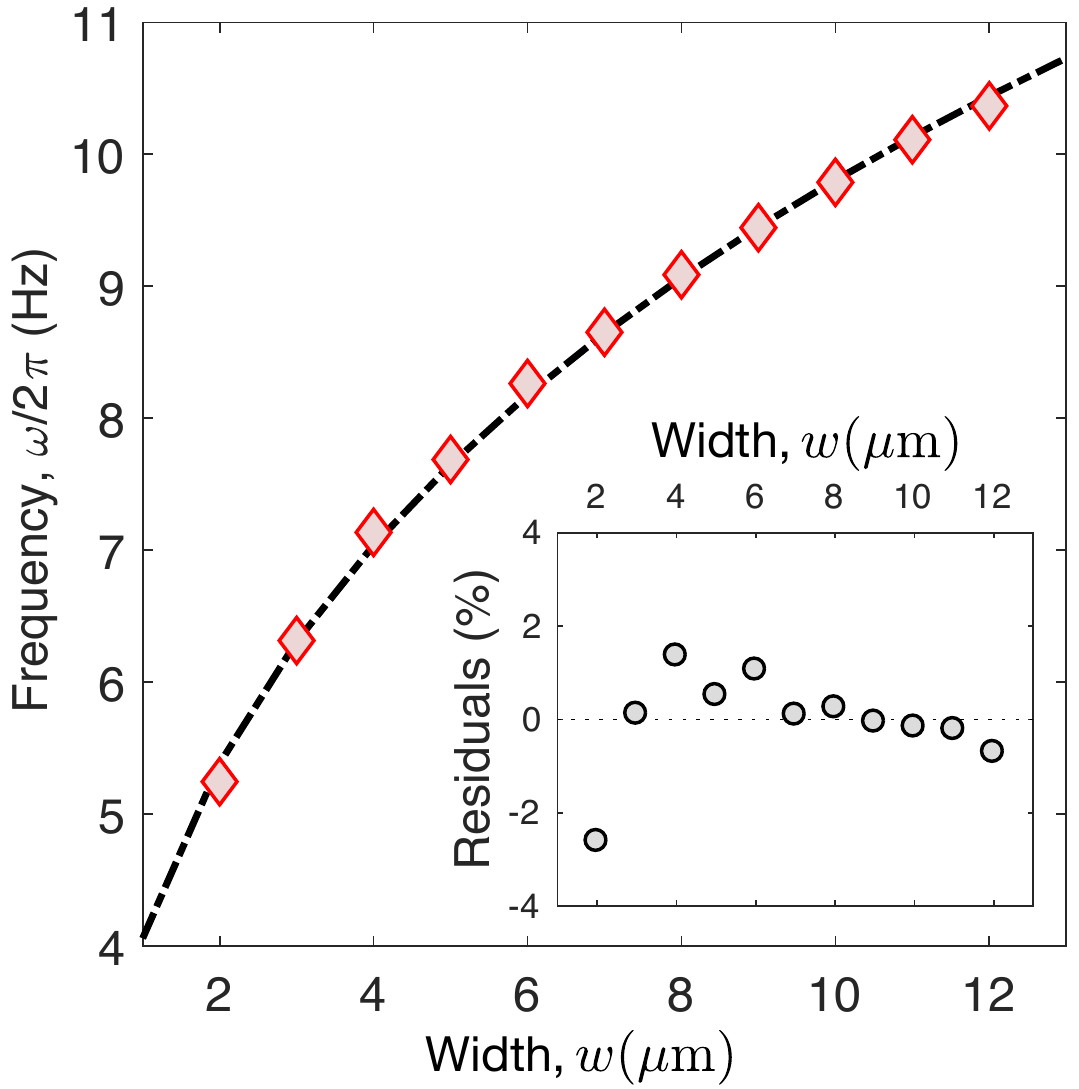}
\caption{Frequency dependence of the superfluid oscillations from GPE simulations as a function of channel width.  The channel has a fixed length $\ell = 10$~$\mu$m and an initial imbalance of $\eta=0.02$ (red diamonds). The fit from the acoustic model (black dashed line) gives $\delta = 2.26(2) \sqrt{\mathcal{S}}$, where $\mathcal{S} = 2 l_z w$ and $l_z = \sqrt{2\mu/m\omega_z^2} \approx 2.36$~$\mu$m. The speed of sound extracted from the averaged 3D density ($c = \sqrt{\bar{n} g /m}$) in the GPE wave function gives $c = 2469$~$\mu$m/s (this is averaged both spatially and over the variation across different channel widths). The density was averaged over all points which were $\geq5\%$  of the maximum density. The value is relatively insensitive to the region chosen; averaging over the region where $n \geq 0.1n_0$ instead gives $c = 2485$~$\mu$m/s. The inset shows the fit residuals.}
\label{fig:varyW_longerL}
\end{figure}
\subsection*{Further testing of the acoustic model}
We have performed further 3D GPE simulations to check the frequency dependence of the superfluid oscillations against the predictions of the acoustic model. Instead of a channel length $\ell = 1.5$~$\mu$m used for the frequency vs. channel width data presented in Fig.~1(d) of the main text, we simulated a system with $\ell = 10$~$\mu$m and initial imbalance $\eta=0.02$, with all other parameters being the same. The results are shown in Fig.~\ref{fig:varyW_longerL}, along with the corresponding fit from the acoustic model, and are in excellent agreement.

\subsection*{Chemical capacitance}
\label{sec:AppendixA}
Instead of using an acoustic model, it is equally valid to consider the potential energy stored in the circuit in terms of a ``chemical" capacitance $C_\mu$, where changes in the chemical potential $\delta \mu$ play the role of the voltage $\delta V$ as considered in Ref.~\cite{eckel_contact_2016}. We may write
\begin{equation}
\frac{1}{2} C_{\mu}\,(\delta\mu)^2 = \frac{1}{2}\int d\mathbf{r}\,  \kappa\, [\delta p(\mathbf{r})]^2 \    = \frac{g}{2} \int d\mathbf{r}\,  [\delta n(\mathbf{r})]^2\  = \delta E_{\rm int}.
\label{eqn:CapacitanceCorrespondence}
\end{equation}
If we assume a spatially uniform redistribution of atoms in each reservoir, $\delta n(\mathbf{r}) = \delta n$, then $\delta \mu = g\, \delta n$ and this leads to the simple expression
\begin{equation}
C_\mu = \frac{\delta N}{\delta \mu} = \frac{\mathcal{V}}{g}.
\label{eqn:SimpleCapacitance}
\end{equation}
Remarkably, the result of \eqnreft{eqn:SimpleCapacitance} can still be obtained even when accounting for the nonuniformity of the background density in $z$. Ignoring weak radial contributions from the optical dipole sheet, the potential is
\begin{align}
V(\mathbf{r})   &= V_{\textrm{sheet}} \exp\left( -2 \frac{z^2}{\sigma_z^2}\right) + V_{\textrm{DMD}} \Theta(x,y), \notag \\
                &\approx \frac{1}{2}m\omega^2_z z^2 + V_{\textrm{DMD}} \Theta(x,y),
\label{eq:Dumbbell_Potential}
\end{align}
where $\Theta(x,y)$ is the binary DMD pattern, $V_{\textrm{DMD}} > \mu$, and $\omega_z = \sqrt{4V_{\textrm{sheet}}/m\sigma_z}$. Applying the  Thomas-Fermi approximation gives $\psi(\mathbf{r}) = \sqrt{[\mu - V(\mathbf{r})]/g}$ for $V(\mathbf{r}) < \mu $ and  $\psi(\mathbf{r})=0$ otherwise.  Within this approximation \eqnreft{eq:Dumbbell_Potential} allows $\psi(\mathbf{r})$  to be more succinctly written as
\begin{equation}
\psi(\mathbf{r}) =  n_0^{1/2} \Theta(x,y) \left(1 - \frac{z^2}{l_z^2}\right)^{1/2},
\end{equation}
where $l_z = \sqrt{2\mu/m\omega_z^2}$ is the Thomas-Fermi radius in $z$. Integrating $|\psi(\mathbf{r})|^2$ hence yields the atom number $N = n_0 A (4l_z /3) = n_0 \mathcal{V}_{\rm{eff}}$ and the relation
 \begin{equation}
\mu = n_0 g =  \frac{1}{2} \left[\frac{3N g(m\omega_z^2)^{1/2} }{2A} \right]^{2/3},
\label{eqn:chemicalPotential}
\end{equation}
where $A$ is the total area of the potential. Following Ref.~\cite{eckel_contact_2016}, assuming a small increment of the atom number $\delta N$, a binomial expansion gives
\begin{align}
\delta \mu  &= \beta \left[ (N + \delta N)^{2/3} -   (N - \delta N)^{2/3}\right] \notag \\               &\approx (\beta N^{2/3}) \frac{4\,\delta N}{3N} = \frac{4 \mu\, \delta N}{3N},     
\end{align}
where $\beta$ represents the constant prefactors in~\eqnreft{eqn:chemicalPotential}. Combining this with \eqnreft{eqn:SimpleCapacitance} leads to an expression for the capacitance:
\begin{equation}
    C_\mu  = \frac{\delta N}{\delta \mu}\approx \frac{3N}{4\mu}.
    \label{eqn:simpleCapacitance}
\end{equation}
 Recalling  $N = n_0 A (4 l_z/3)$ and $\mu = n_0 g$, this reduces to \eqnreft{eqn:CapacitanceCorrespondence}, identical to the uniform system.\\
\begin{figure}[!t]
\centering
\includegraphics[width =.85\columnwidth]{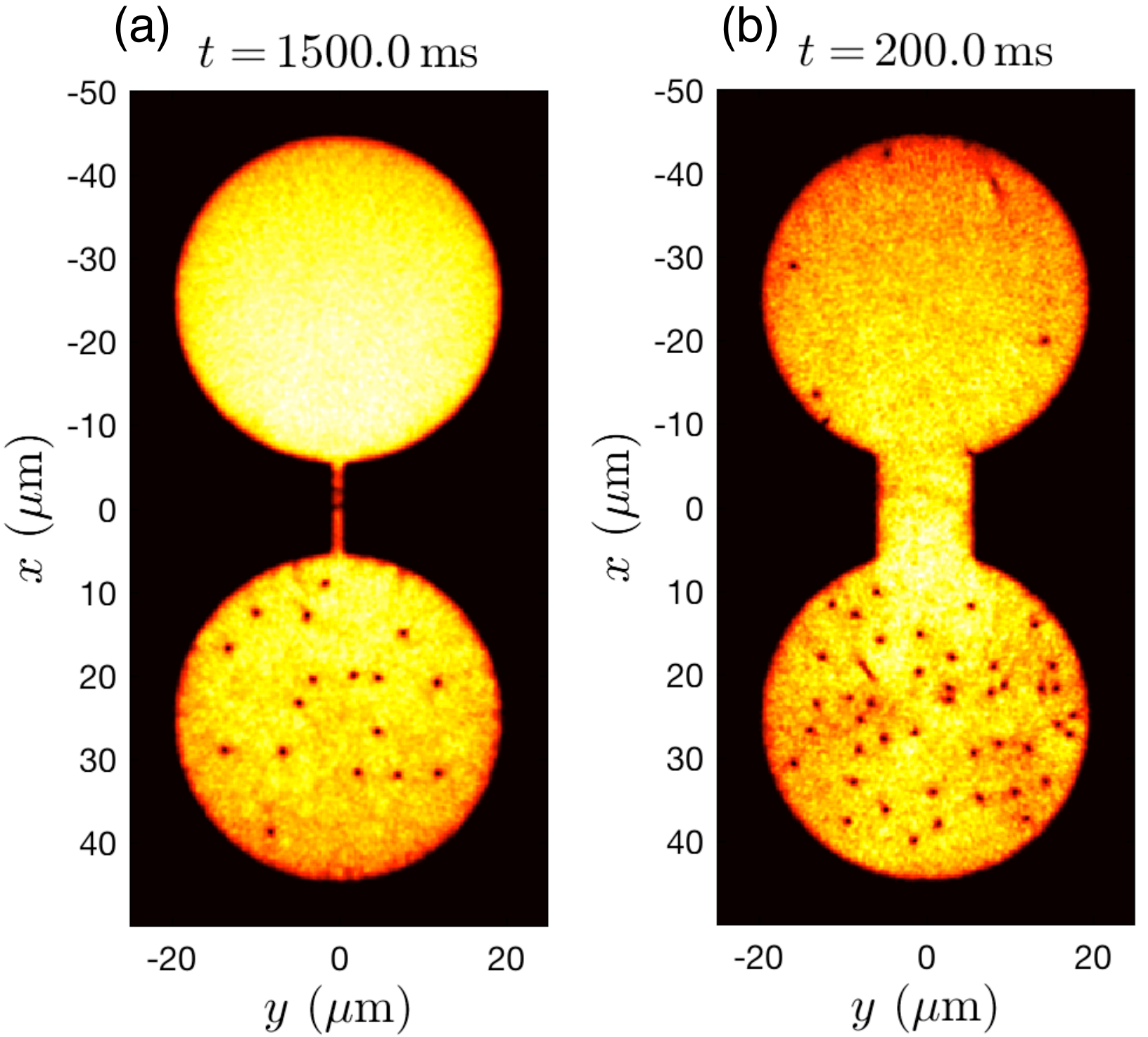}
\caption{(a) From Supplemental Movie 1:  Snapshot of the \emph{in situ} density from a GPE simulation illustrating the dissipative circuit dynamics for a narrow channel of width $w = 2.0$~$\mu$m and length $\ell = 10$~$\mu$m. The circuit is initially biased to $\eta = 1$ with $N = 7.5 \times 10^5$ atoms. The simulation shows that vortices are created as the initially empty reservoir is filled. (b) From Supplemental Movie 2: Snapshot of the \emph{in situ} density from a GPE simulation illustrating the dissipative circuit dynamics for a wide channel of width $w = 12.5$~$\mu$m  and  length $\ell = 10$~$\mu$m. The circuit is initially biased to $\eta = 1$ with $N = 7.5 \times 10^5$ atoms. The dynamics produce considerably more vortices compared to the narrow channel width case.}
\label{fig:mov1+2}
\end{figure}

\subsection*{Calculation of compressible and incompressible energies}
The total kinetic energy of a superfluid can be decomposed into~\cite{Nore:1997, Bradley:2012}
\begin{equation}
    E_K = E_c + E_i + E_Q,
\end{equation}
where $E_c$ and $E_i$ are the compressible and incompressible energies, respectively, and $E_Q$ is the quantum pressure term. Given a GPE wave function $\psi(\textbf{r})$ with density $n(\textbf{r}) = |\psi(\textbf{r})|^2$ and velocity field $\textbf{u}(\textbf{r}) = (\hbar/m) \textrm{Im}\left\{ \psi^*(\textbf{r}) \nabla^2 \psi(\textbf{r}) \right\} /|\psi(\textbf{r})|^2$, we can compute these energy contributions via
\begin{align}
    E_c     &= \frac{1}{2}m \int d\textbf{k}\, \frac{\textbf{k} \cdot \tilde{ \textbf{w}}(\textbf{k})}{|\textbf{k}|^2}, \label{Eqn_E_c}\\
    E_i     &= \frac{1}{2}m \int d\textbf{k}\, |\tilde{\textbf{w}}(\textbf{k})|^2 - E_c, \label{Eqn_E_i} \\
    E_Q     &= \frac{\hbar^2}{2m}\int d\textbf{k}\, |\textbf{k}|^2 \left| \mathcal{F}\left[ \sqrt{n} \right](\textbf{k}) \right|^2,
\end{align}
where $\tilde{\textbf{w}}(\textbf{k})$ and $\mathcal{F}\left[ \sqrt{n} \right](\textbf{k})$ are the Fourier transforms of $\textbf{w}(\textbf{r}) = \sqrt{n(\textbf{r})}\textbf{u}(\textbf{r})$ and $\sqrt{n(\textbf{r})}$, respectively. Equations~(\ref{Eqn_E_c}) and (\ref{Eqn_E_i}) were used to compute the fractions of compressible to incompressible energy shown in Fig.~4 of the main text.

\subsection*{Supplemental movies}
Figure~\ref{fig:mov1+2} shows snapshots of the superfluid density from GPE simulations of the experiments in the dissipative regime.\\

\noindent\textbf{Supplemental Movie 1:}
Illustrates the dissipative circuit dynamics of the \emph{in situ} density from a GPE simulation for a narrow channel with width $w = 2.0$~$\mu$m and length $\ell = 10$~$\mu$m. The circuit is initially biased to $\eta = 1$ with $N = 7.5 \times 10^5$ atoms. The simulation shows that vortices are created as the initially empty reservoir is filled.\\

\noindent\textbf{Supplemental Movie 2:}  Illustrates the dissipative circuit dynamics of the \emph{in situ} density from a GPE simulation for a wide channel with width $w = 12.5$~$\mu$m and length $\ell = 10$~$\mu$m. The circuit is initially biased to $\eta = 1$ with $N = 7.5 \times 10^5$ atoms. The dynamics produce considerably more vortices compared to the narrow channel width case.

\end{document}